\documentclass[preprint]{aastex631}

\usepackage{soul}     % for underlining over line breaks
\usepackage{amsmath}  % additional math formating
\usepackage{amssymb}  % additional math symbols
\usepackage{MnSymbol} % additional math symbols
\usepackage{wasysym}  % additional math symbols
\usepackage{multirow} % multirow table package
\usepackage{xcolor,colortbl}  % color package
\definecolor{Gray}{gray}{0.9} % create color for table highlights

\begin{document}

\title{Energy and temperature dependencies for electron-induced sputtering from H$_2$O-ice: Implications for the icy Galilean moons}

\correspondingauthor{Rebecca Carmack}
\email{rcarmack@nau.edu}

\author[0000-0002-3306-3206]{Rebecca A. Carmack}
\affiliation{Department of Astronomy and Planetary Science, Northern Arizona University, Box 6010, Flagstaff, AZ 86011, USA}

\author[0000-0002-1551-3197]{Mark J. Loeffler}
\affiliation{Department of Astronomy and Planetary Science, Northern Arizona University, Box 6010, Flagstaff, AZ 86011, USA}
\affiliation{Center for Materials Interfaces in Research and Applications, Northern Arizona University, Flagstaff, AZ 86011, USA}

\begin{abstract}

To better assess the role that electrons play in exosphere production on icy-rich bodies, we measured the total and O$_2$ sputtering yields from H$_2$O-ice for electrons with energies between 0.75 and 10 keV and temperatures between 15 and 124.5 K. We find that both total and O$_2$ yields increase with decreasing energy over our studied range, increase rapidly at temperatures above 60 K, and that the relative amount of H$_2$O in the sputtered flux decreases quickly with increasing energy. Combining our data with other electron data in literature, we show that the accuracy of a widely used sputtering model can be improved significantly for electrons by adjusting some of the intrinsic parameter values. Applying our results to Europa, we estimate that electrons contribute to the production of the O$_2$ exosphere equally to all ion types combined. In contrast, sputtering of O$_2$ from Ganymede and Callisto appears to be dominated by irradiating ions, though electrons still likely contribute a non-negligible amount. While our estimates could be further refined by examining the importance of spatial variations in electron flux, we conclude that, at the very least, electrons seem to be important for exosphere production on icy surfaces and should be included in future modeling efforts.

\end{abstract}


\section{Introduction} \label{sec:intro}

Planetary bodies in our solar system that lack protection from a significant atmosphere are subjected to a number of irradiating particles, such as ions, electrons, photons, and cosmic rays. These particles alter the surface composition and/or structure, as well as eject surface material in a process known as sputtering. The sputtering of surface material can create surface bound exospheres on both rocky \citep{stern_lunar_1999, wurz_lunar_2007, wurz_self-consistent_2010, gamborino_mercurys_2019} and icy bodies \citep{hall_detection_1995,ip_energetic_1997, cunningham_detection_2015, ligier_surface_2019, carnielli_simulations_2020, liuzzo_variability_2020, plainaki_kinetic_2020, paranicas_energetic_2022, carberry_mogan_callistos_2023, de_kleer_optical_2023}.

\cite{hall_detection_1995} identified an exosphere on Europa containing atomic oxygen and hypothesized that incoming energetic particles cause the dissociation and excitation of molecular O$_2$ in the atmosphere, which in turn is predicted to be sputtered off Europa's icy surface along with molecular hydrogen and H$_2$O \citep{cunningham_detection_2015}. Since atomic and molecular hydrogen are light enough to dissipate into space and H$_2$O falls back onto the surface, the main component of Europa's exosphere is oxygen \citep{johnson_planetary_1982, johnson_composition_2009}. Similar sputtering processes may occur on Ganymede \citep{ligier_surface_2019, paranicas_energetic_2022} and Callisto \citep{cunningham_detection_2015, carberry_mogan_callistos_2023}, although the interactions of irradiating particles with those surfaces are more complex.

While both ions and electrons can cause sputtering from icy surfaces, ions have been the main focus of previous experimental (see \citeauthor{baragiola_sputtering_2003} \citeyear{baragiola_sputtering_2003} and \citeauthor{teolis_water_2017} \citeyear{teolis_water_2017} for a summary) and sputtering/exosphere modeling studies \citep{fama_sputtering_2008, cassidy_magnetospheric_2013, teolis_water_2017, addison_effect_2022, pontoni_simulations_2022}. The lack of prior attention to electrons is at least partially due to early laboratory data showing that the sputtering yield ($Y$; the average number of molecules removed from a target material per incident particle) for a single 100 keV electron \citep{heide_observations_1984} is 1000 to 10,000 times lower than the sputtering yield for a hydrogen or oxygen ion at similar energies \citep{shi_sputtering_1995}. However, this difference in sputtering yields may not be that extreme, as the stopping cross section, a parameter which correlates with sputtering, is very low for 100 keV electrons and increases with decreasing energy until it peaks near 0.12 keV \citep{castillo-rico_stopping_2021}. Regardless, electrons contribute $\sim$90$\%$ of particles and $\sim$80$\%$ of total energy measured near Europa, and smaller but still significant portions of particles/energy measured near Ganymede and Callisto \citep{cooper_energetic_2001}. The large flux of electrons near these icy moons could make them important for exosphere production even if electrons are individually less efficient at sputtering than ions.

Previous experiments irradiating H$_2$O-ice with very low-energy (5 to 100 eV) electrons found that sputtering occurs for energies greater than $\sim$10 eV \citep{sieger_production_1998, orlando_role_2003}, and that O$_2$ sputtering yields increase with increasing electron energy between $\sim$10 and 100 eV \citep{sieger_production_1998, orlando_role_2003}, remain relatively constant at low temperatures ($\lesssim$80 K; \citeauthor{petrik_electron-stimulated_2005} \citeyear{petrik_electron-stimulated_2005}; \citeauthor{davis_contribution_2021} \citeyear{davis_contribution_2021}), and increase with increasing temperature above 80 K \citep{sieger_production_1998, orlando_role_2003, petrik_electron-stimulated_2005, davis_contribution_2021}. 

Three groups have investigated the composition of material sputtered by higher electron energies \citep{abdulgalil_electron-promoted_2017, galli_02_2018, davis_contribution_2021}. Both \cite{abdulgalil_electron-promoted_2017} and \cite{galli_02_2018} detected little to no H$_2$O sputtered by 0.2 to 10 keV electrons near 100 K, while our group \citep{davis_contribution_2021} determined H$_2$O dominates material sputtered by 0.5 keV electrons at low temperatures ($\leq$60 K) and constitutes $\sim$1/5 of sputtered molecules at 100 K. Whether these differences between laboratory groups are mainly a consequence of the composition of sputtered material depending on electron energy, as has been observed for ions \citep{brown_electronic_1984, bar-nun_ejection_1985, baragiola_atomic_2002}, or due to other factors is currently unclear.

Quantifying the composition of material sputtered from H$_2$O-ice as a function of electron energy and temperature is critical to properly model sputtering rates and exosphere production on icy bodies. Recently, we estimated Europa's global production of O$_2$ due to electrons by combining our laboratory data \citep{davis_contribution_2021} with the scaled down ion sputtering model from \cite{teolis_water_2017}. We found that electrons could be responsible for sputtering as much or more O$_2$ as all incoming ions combined \citep{davis_contribution_2021}. However, due to a lack of experimental data, we assumed that the composition of sputtered material did not change with electron energy in our calculation. 

Thus, here we measure the composition of the sputtering yield as a function of both electron energy and irradiation temperature, using a combination of microbalance gravimetry and mass spectrometry. With our new data, we use Markov chain Monte Carlo methods to determine electron versions of intrinsic model values that \cite{teolis_water_2017} determined for ions. Lastly, we use our optimized electron sputtering model to recalculate our previous estimate of the global production rate of O$_2$ by electrons irradiating Europa \citep{davis_contribution_2021} and compare our updated model to additional estimates in literature for sputtering of O$_2$ from Europa, Ganymede, and Callisto, allowing us to better assess the role of electrons in icy satellite exosphere production. 

\section{Experimental Methods} \label{sec:style}

We performed all experiments within a stainless steel ultra-high vacuum chamber at a base pressure of $\sim$3 x 10$^{-9}$ Torr \citep{meier_sputtering_2020, davis_contribution_2021}. We estimate that the pressure near the sample is 10 to 100 times lower due to a thermal-radiation shield in place around the sample. An Inficon IC6 quartz-crystal microbalance (QCM) with an optically flat gold mirror electrode served as the sample substrate and is mounted onto a rotatable closed-cycle helium cryostat centered inside of the experimental chamber. The cryostat is capable of maintaining temperatures between $\sim$10 and 300 K. 

We prepared H$_2$O (HPLC grade) samples in a separate manifold attached to the chamber and grew samples at 100 K at near normal incidence with a deposition rate of $\sim$2 x 10$^{15}$ H$_2$O cm$^{-2}$ s$^{-1}$ to an average column density of (5.4$^{+1.3}_{-0.4}$) x 10$^{18}$ H$_2$O cm$^{-2}$ ($\sim$2 $\mu$m), with the error representing the full range of column densities used in this study. The resulting sample thickness is sufficient to avoid any enhancement in our measured yields for all electron energies studied here \citep{meier_sputtering_2020}. We grew fresh films for all electron energies and irradiation temperatures reported here, since sample irradiation history can affect sputtering yields \citep{meier_sputtering_2020}. After growth, we changed the sample temperature to the irradiation temperature of interest (between 14 and 125 K). The lower limit ensured we could consistently stabilize the temperature and the higher limit is below the temperature ($\sim$130 K) where H$_2$O begins to sublimate \citep{sack_sublimation_1993} and out diffusion of radiolytically O$_2$ produced below the near-surface becomes important \citep{teolis_mechanisms_2005}.

We irradiated the sample with an EGG-3103C Kimball Physics electron gun at an incident angle of 12.5$^{\circ}$ with respect to the surface normal with 0.75 to 10 keV electrons. In all experiments, we rastered the beam in an approximately 1 x 1 cm square, which is larger than the exposed surface of our QCM ($\sim$8 mm diameter). We measured the electron flux before and after irradiation using a retractable Faraday cup to be (2.7$\pm$0.8) x 10$^{13}$ electrons cm$^{-2}$ s$^{-1}$. During irradiation, the flux varied by $\lesssim$2$\%$ for all energies except for 10 keV which varied up to $\sim$9$\%$. We analyzed any gases present in the chamber, including residual background and material sputtered from the sample during irradiation, using an Ametek Dymaxion Mass Spectrometer (DYMAX-100) aligned 12.5$^{\circ}$ from the sample normal opposite the electron gun. After each irradiation, we desorbed our ice by turning off the cryostat and allowing the substrate to return to $\sim$300 K overnight. In our analysis, we include previous work done by our group in \cite{davis_contribution_2021} with 0.5 keV electrons, as they used the same setup and approach as we do here. 

\section{Results} \label{sec:results}

\begin{figure}[t]
\centering
\includegraphics[width=0.5\textwidth]{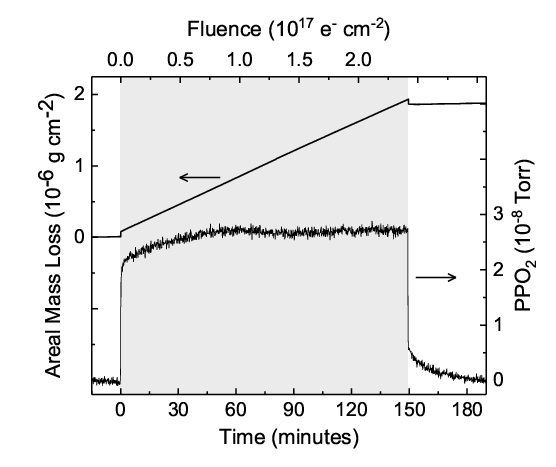} 
\caption{Areal mass loss during irradiation (top) measured by the QCM and O$_2$ partial pressure (PPO$_2$, bottom) measured by the mass spectrometer for a H$_2$O-ice sample at 115 K irradiated with 1 keV electrons. The highlighted area shows when the electron beam was irradiating the sample.}
\label{fig:raw_data}
\end{figure}

During irradiation, we see a clear increase in the partial pressure of H$_2$O, O$_2$, and H$_2$ for each electron energy and temperature studied. However, the background signals for H$_2$O and H$_2$ are 1 to 2 orders of magnitude higher than the background for O$_2$, and therefore are highly affected by baseline changes. Additionally, the cooled thermal-radiation shield around our sample acts as a potential cold trap for H$_2$O but is less likely to trap more volatile species like O$_2$ and H$_2$ \citep{davis_contribution_2021}. Because of these barriers to accurately interpreting our partial pressure signals for H$_2$O and H$_2$, we only consider the partial pressure signal for O$_2$ (PPO$_2$) in our data analysis. 

Figure \ref{fig:raw_data} shows the areal mass loss as monitored by the QCM alongside the baseline subtracted PPO$_2$ for a sample irradiated with 1 keV electrons at 115 K, with the highlighted area showing when the electron beam was irradiating the sample. When irradiation begins, there is an initial period when the PPO$_2$ rises until it reaches a peak, after which it levels out at equilibrium for the remainder of the irradiation. In cases where we see a peak ($\gtrsim$115 K), the fluence required to reach the peak and subsequent equilibrium is energy and temperature dependent. However, all experiments reached equilibrium between fluences of $\sim$(0.4 – 3) x 10$^{17}$ electrons cm$^{-2}$. Generally, we used the equilibrium value to determine the PPO$_2$, but in cases where we observed a peak we took the average of the peak and equilibrium values. We incorporated the differences between the peak and equilibrium PPO$_2$ values into our error. Regardless of irradiation temperature, when the electron gun is blocked the PPO$_2$ takes time to return to zero. This could be due to any sputtered O$_2$ remaining in the chamber slowly being pumped out of our system.

\begin{figure}[t]
\centering
\includegraphics[width=0.525\textwidth]{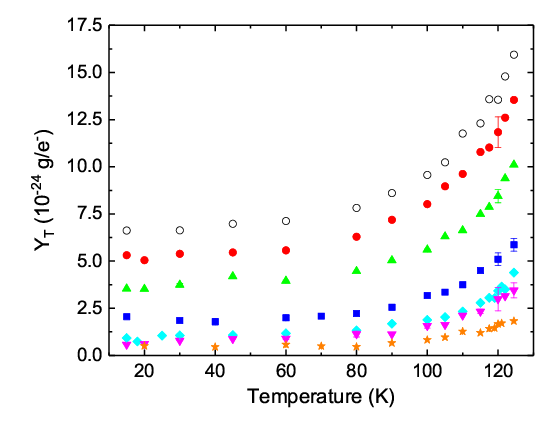}
\caption{Total sputtering yield ($Y_T$) versus irradiation temperature for 0.5 ($\circ$; from \citeauthor{davis_contribution_2021} \citeyear{davis_contribution_2021}), 0.75 (\textcolor{red}{$\bullet$}), 1 (\textcolor{green}{$\filledtriangleup$}), 2 (\textcolor{blue}{$\blacksquare$}), 4 (\textcolor{cyan}{$\blacklozenge$}), 6 (\textcolor{magenta}{$\blacktriangledown$}), and 10 (\textcolor{orange}{$\bigstar$}) keV electrons.}
\label{fig:YT_v_T}
\end{figure} 
\begin{figure}[b]
\centering
\includegraphics[width=0.525\textwidth]{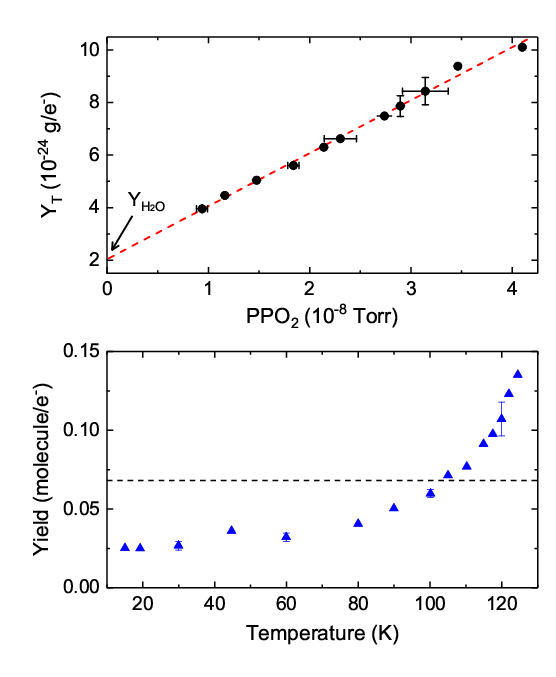}
\caption{Top: Total sputtering yield ($Y_T$) versus O$_2$ partial pressure (PPO$_2$) for samples irradiated with 1 keV electrons and the resulting linear fit (red dashed line). When the PPO$_2$ is zero, the only material sputtered is H$_2$O. Bottom: H$_2$O (dashed line) and O$_2$ (\textcolor{blue}{$\filledtriangleup$}) molecular sputtering yields versus irradiation temperature for 1 keV electrons. The molecular yield of H$_2$ is presumed to be twice that of O$_2$.}
\label{fig:1keVdata}
\end{figure}

Figure \ref{fig:YT_v_T} shows the total sputtering yield ($Y_{T}$; in terms of the sample's total mass loss) for 0.5 to 10 keV electrons at temperatures between 14 and 124.5 K. $Y_{T}$ is approximately constant below 60 K for all energies, although electron energies below $\leq$2 keV show a small ($\lesssim$10$\%$) increase in $Y_T$ between 15 and 60 K. For higher energies, we observe a similar trend but cannot say definitively due to increased variation in $Y_T$. Above 60 K, $Y_{T}$ clearly increases with temperature. For example, $Y_{T}$ increases for all energies by a factor of $\sim$1.5 between 60 and 100 K, and a factor of 2 to 3 between 60 and 120 K.
\begin{deluxetable*}{cc|D@{$\pm$}DD@{$\pm$}DD@{$\pm$}DD@{$\pm$}D}[t]
\tablecaption{Summary of Laboratory Sputtering Yields \label{Table:Summary Results}}
\tablehead{
\colhead{\textbf{Energy}} & \colhead{\textbf{Temperature}} & \multicolumn4c{\textbf{Total Yield\tablenotemark{a}}} & \multicolumn4c{\textbf{H$_2$O Yield}} & \multicolumn4c{\textbf{H$_2$O Yield}} & \multicolumn4c{\textbf{O$_2$ Yield\tablenotemark{b}\tablenotemark{c}}} \\
\colhead{(keV)} & \colhead{(K)} & \multicolumn4c{(10$^{-24}$ g/e$^-$)} & \multicolumn4c{(10$^{-24}$ g/e$^-$)} & \multicolumn4c{(H$_2$O/e$^-$)} & \multicolumn4c{(O$_2$/e$^-$)}}
\decimals
\startdata
\multirow{3}{*}{0.5\tablenotemark{d}} & 60  & 7.13 & 0.4 & 5.1 & 0.3 & 0.17 & 0.01 & 0.034 & 0.01 \\
& 100 & 9.57 & 0.2 & 5.1 & 0.3 & 0.17 & 0.01 & 0.075 & 0.01 \\
& 120 & 13.56 & 0.4 & 5.1 & 0.3 & 0.17 & 0.01 & 0.141 & 0.02 \\ \hline
\multirow{3}{*}{0.75} & 60 & 5.57 & 0.01 & 2.8 & 0.2 & 0.094 & 0.007 & 0.046 & 0.01 \\
& 100 & 8.02 & 0.03 & 2.8 & 0.2 & 0.094 & 0.007 & 0.087 & 0.02 \\
& 120 & 11.84 & 0.81 & 2.8 & 0.2 & 0.094 & 0.007 & 0.151 & 0.02 \\ \hline
\multirow{3}{*}{1} & 60 & 3.96 & 0.03 & 2.0 & 0.1 & 0.068 & 0.004 & 0.032 & 0.003 \\
& 100 & 5.61 & 0.03 & 2.0 & 0.1 & 0.068 & 0.004 & 0.060 & 0.003 \\
& 120 & 8.44 & 0.52 & 2.0 & 0.1 & 0.068 & 0.004 & 0.107 & 0.010 \\ \hline
\multirow{3}{*}{2} & 60 & 1.99 & 0.1 & 0.48 & 0.24 & 0.016 & 0.008 & 0.025 & 0.006 \\
& 100 & 3.17 & ... & 0.48 & 0.24 & 0.016 & 0.008 & 0.045 & 0.004 \\
& 120 & 5.10 & 0.3 & 0.48 & 0.24 & 0.016 & 0.008 & 0.077 & 0.096 \\ \hline
\multirow{3}{*}{4} & 60 & 1.17 & 0.04 & 0.18 & 0.45 & 0.006 & 0.015 & 0.017 & 0.008 \\
& 100 & 1.89 & 0.04 & 0.18 & 0.45 & 0.006 & 0.015 & 0.029 & 0.008 \\
& 120 & 3.42 & 0.17 & 0.18 & 0.45 & 0.006 & 0.015 & 0.054 & 0.011 \\ \hline
\multirow{3}{*}{6\tablenotemark{e}}& 60 & 0.90 & 0.06 & -0.29 & 0.3 & 0.0 & 0.01 & 0.015 & 0.006 \\
& 100 & 1.58 & 0.12 & -0.29 & 0.3 & 0.0 & 0.01 & 0.026 & 0.007 \\
& 120 & 2.99 & 0.62 & -0.29 & 0.3 & 0.0 & 0.01 & 0.050 & 0.015 \\ \hline
\multirow{3}{*}{10} & 60 & 0.58 & 0.20 & 0.12 & 0.22 & 0.004 & 0.007 & 0.008 & 0.007 \\
& 100 & 0.82 & ... & 0.12 & 0.22 & 0.004 & 0.007 & 0.012 & 0.008 \\
& 120 & 1.67 & 0.25 & 0.12 & 0.22 & 0.004 & 0.007 & 0.026 & 0.011
\enddata
\tablenotetext{a}{error from the spread in values from repeated experiments}
\tablenotetext{b}{molecular yield of H$_2$ is twice that of O$_2$ ($Y_{H_2}=2*Y_{O_2}$)}
\tablenotetext{c}{error in O$_2$ estimated from either propagating the error from the repeatability of experiments or from the difference between our measured $Y_{O_2}$ and the linear trend for $Y_{O_2}$ versus $S_e$ for a given temperature (shown in Figure \ref{fig:mass yields} for 60 K), whichever is larger}
\tablenotetext{d}{from \cite{davis_contribution_2021}}
\tablenotetext{e}{the y-intercept of $Y_T$ versus the $PPO_2$ for 6 keV is negative but zero within error, so we assume the amount of water sputtered is zero}
\end{deluxetable*}

In order to determine the composition of sputtered material, we use the same approach described in \cite{davis_contribution_2021} for each electron energy studied. We assume the amount of H$_2$O sputtered from ice for a given electron energy is constant with temperature, previously shown to be true for temperatures $\lesssim$130 K \citep{boring_ion-induced_1983, petrik_electron-stimulated_2005}. Figure \ref{fig:1keVdata} (top) shows $Y_{T}$ versus the PPO$_2$ for all experiments where we irradiated our sample with 1 keV electrons above 60 K. Each data point is an experiment completed at a different temperature (if a temperature was repeated more than once, the average data point for that temperature is shown). We do not include experiments performed at temperatures below 60 K in the analysis, because baseline variations in the mass spectrometer signal occur more frequently at lower temperatures and because $Y_T$ is $\sim$constant below 60 K. We calculate the sputtering yield of H$_2$O ($Y_{H_2O}$) for a given electron energy by extrapolating the PPO$_2$ to zero (i.e. the y-intercept in the top of Figure \ref{fig:1keVdata}), implying no O$_2$ (or consequentially H$_2$) is sputtered from the sample. We tested whether the y-intercept was unique for a given electron energy by repeating a suite of experiments under the same conditions (energy, temperatures, etc.) but using a different mass spectrometer multiplier voltage. In those experiments, we find that the data remains linear (although the slope changes) and the y-intercept remains the same.

The difference between $Y_T$ and $Y_{H_2O}$ gives the portion of sputtered material that is comprised of radiolytic products O$_2$ and H$_2$. The sputtering yields for O$_2$ ($Y_{O_2}$) and H$_2$ ($Y_{H_2}$) are then differentiated from each other by multiplying the sputtered mass of radiolytic products by the mass fraction of O$_2$ and H$_2$ in the relation $2\cdot H_2O \rightarrow O_2 + 2\cdot H_2$ \citep{brown_linear_1980}. We show the compositional breakdown of molecules sputtered by 1 keV electrons in Figure \ref{fig:1keVdata} (bottom) for each irradiation temperature studied. While we did not measure $Y_{H_2}$ directly, we assume it is twice that of $Y_{O_2}$ (see above). 

We apply the same analysis to 0.75, 2, 4, 6, and 10 keV electrons and provide a sampling of representative total mass yields, H$_2$O mass yields and H$_2$O and O$_2$ molecular yields in Table \ref{Table:Summary Results} (for the entirety of our data see Table \ref{Table: Results} in Appendix \ref{Appendix: Data}). We find that the composition of sputtered material varies strongly across 0.5 to 10 keV and 60 to 125 K. At low temperatures ($\leq$60 K), H$_2$O makes up as much as 65$\%$ of the sputtered flux for 0.5 keV electrons, about 40$\%$ for 1 keV electrons, but only comprises about 20$\%$ for 2 keV electrons. Above 4 keV, the contribution of H$_2$O to the sputtered flux is essentially zero within our error. At higher temperatures ($>$60 K), H$_2$O yields trend similarly with energy as they do at low temperatures, however the relative contribution of H$_2$O at each temperature is lower due to the increased production of radiolytic O$_2$ and H$_2$. For instance, at 120 K H$_2$O makes up about 30$\%$ of the sputtered flux at 0.5 keV, about 20$\%$ at 1 keV, but drops to about 6$\%$ of the flux at 2 keV. 

\begin{figure}[b]
\centering
\includegraphics[width=0.5\textwidth]{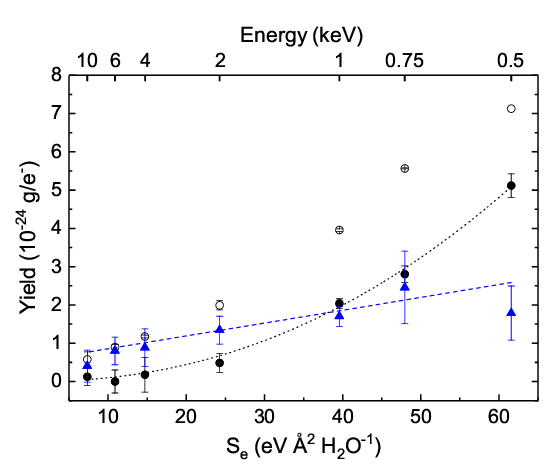}
\caption{Total ($\circ$), H$_2$O ($\bullet$), and O$_2$ (\textcolor{blue}{$\blacktriangle$}) mass yields of sputtered molecules versus electron stopping cross section ($S_e$) for all electron energies (0.5 keV from \citeauthor{davis_contribution_2021} \citeyear{davis_contribution_2021}, the rest from this work) at an irradiation temperature of 60 K. We also plot the linear fit to O$_2$ yields (blue dashed line) and the nearly quadratic fit to H$_2$O yields (black dotted line). The fit for H$_2$O is applicable for all temperatures, and follows $y=a(S_e)^n$ where $a$=6.65 x 10$^{-28}$ and $n$=2.17.}
\label{fig:mass yields}
\end{figure}

Figure \ref{fig:mass yields} shows the total, H$_2$O, and O$_2$ mass yields versus electron stopping cross section ($S_e$) for irradiation at 60 K. $Y_{H_2O}$ (in g/e$^-$, for all temperatures) is nearly quadratic with $S_e$, and well fit to $y=a(S_e)^n$ where $a$=6.65 x 10$^{-28}$ and $n$=2.17. At lower temperatures, the total sputtering yield is superlinearly related to $S_e$, which is consistent with previous studies \citep{meier_sputtering_2020}. As temperature increases, the trend of $Y_T$ with $S_e$ becomes more linear, likely because $Y_{O_2}$ (and $Y_{H_2}$) appear to increase linearly with $S_e$ for all temperatures, though given the error on O$_2$ yields this is hard to state definitively. 

\section{Comparison to Other Experiments} \label{sec:comparison}

In this study, we expanded on our previous work \citep{meier_sputtering_2020, davis_contribution_2021} to investigate the composition of the sputtering yield as a function of irradiation temperature and electron energy. Below, we compare and discuss our results with previous ion and electron work measuring total sputtering yields, as well as studies that have made estimates of the main species sputtered from H$_2$O-ice.

\subsection{Total Yields} \label{sec:comparison total Y}

Previous work on the sputtering of H$_2$O-ice with light ions at low temperatures ($\lesssim$80 K) found $Y_T$ to be proportional to $S_e$ following a superlinear and in some cases quadratic dependence \citep{brown_energy_1980, brown_linear_1980, shi_sputtering_1995, baragiola_sputtering_2003}. Our group finds a similar superlinear dependence on $Y_T$ with $S_e$ for electron energies between 0.5 and 10 keV irradiating H$_2$O-ice at lower temperatures (this work, \citeauthor{meier_sputtering_2020} \citeyear{meier_sputtering_2020}), but the trend progressively becomes more linear with increasing temperature. The dependence of $Y_T$ with $S_e$ ranging from quadratic to linear is likely due to changes in the composition of the sputtered flux (see Section \ref{sec:comparison compositional Y}). In contrast, \cite{galli_02_2018} found $Y_T$ was independent of $S_e$ between 0.2 and 3 keV for thin films irradiated with electrons at 90 K. We suspect that their observed constancy of $Y_T$ with energy is likely a consequence of using previously irradiated samples, as processed samples can show enhancements in $Y_T$ by a factor of $\sim$3 to 6 at 60 K \citep{meier_sputtering_2020}, which we attribute to the buildup of O$_2$ beneath the sample's surface. 

At low temperatures ($\leq$60 K), we observe a slight ($\lesssim$10$\%$) increase in $Y_T$ between 15 and 60 K, which is consistent with previous studies for electrons \citep{petrik_electron-stimulated_2005, davis_contribution_2021}. For higher temperatures, we find that $Y_T$ increases rapidly above $\sim$60 K for all electron energies studied, consistent with our previous work with 0.5 keV electrons \citep{davis_contribution_2021} and with previous ion irradiation studies \citep{brown_electronic_1984,baragiola_atomic_2002, baragiola_sputtering_2003, fama_sputtering_2008}.

\subsection{Composition of the Sputtered Flux} \label{sec:comparison compositional Y}

We find that the composition of the sputtered flux depends on both electron energy and irradiation temperature. Changes in the composition of our sputtered flux with electron energy are consistent with previously observed experimental trends for ions, which have shown the composition changes with ion energy and ion type. More specifically, experiments with 1.5 MeV He$^+$ found that H$_2$O makes up $\sim$90$\%$ of the sputtered flux at low temperatures \citep{brown_electronic_1984}, while only about $\sim$30$\%$ is H$_2$O for 1 to 5 keV H$^+$ \citep{bar-nun_ejection_1985}. Variations in composition are also observed with heavier ions: samples irradiated with 1 to 5 keV Ne$^+$ found H$_2$O comprises about 60$\%$ of the sputtered flux at 1 keV but only about 30$\%$ at 5 keV. Additionally, studies using 100 keV Ar$^+$ show that $\sim$75$\%$ of the sputtered flux is H$_2$O \citep{baragiola_atomic_2002}.

In addition, we find that H$_2$O, and possibly also O$_2$, has a quantifiable dependence on $S_e$ (Figure \ref{fig:mass yields}). Our observed quadratic dependence for $Y_{H_2O}$ is consistent with what has been seen for $Y_T$ in previous studies with fast ions \citep{brown_linear_1980, baragiola_sputtering_2003}. Given that H$_2$O is the dominant component sputtered by fast ions \citep{brown_electronic_1984}, we speculate that the quadratic dependence for $Y_{H_2O}$ observed in our experiments is also a result of excitation pairs overlapping at the surface \citep{brown_linear_1980, baragiola_sputtering_2003}. For O$_2$, the possible linear relation with $S_e$ suggests that the multiple reactions required to form O$_2$ from H$_2$O \citep{boring_ion-induced_1983, teolis_mechanisms_2005} may occur from a single electron breaking multiple bonds as it travels into the ice.

We can also compare our results to the two other groups who estimate the composition of flux sputtered from H$_2$O-ice by $\sim$keV electrons. \cite{abdulgalil_electron-promoted_2017} irradiated films coated with islands of C$_6$H$_6$ with $\sim$0.25 keV electrons at 112 K. During irradiation, they observed a H$_2$ signal, a much weaker O signal, but no H$_2$O signal above the noise level; no measurement of O$_2$ was reported. They conclude that H$_2$ and O$_2$ are the dominant species removed during irradiation, which is inconsistent with our findings that H$_2$O makes up 45$\%$ of our total sputtering yield at 110 K for 0.5 keV electrons. Interestingly, \cite{galli_02_2018} irradiated several H$_2$O-ice types (thin films, frost, etc.) with 0.2 to 10 keV electrons at $\sim$90 K and, similar to \cite{abdulgalil_electron-promoted_2017}, did not see a rise in H$_2$O above their detection limit while irradiating. They report an average composition between 0.4 and 10 keV for their frost and fine-grained ice samples, estimating the contribution of H$_2$O to the sputtering yield to be $<$10$\%$. Although it is unclear what energies were averaged, this upper limit may be in-line with our findings. For instance, H$_2$O only contributes $\sim$13$\%$ for 2 keV electrons at 90 K and subsequently less at energies approaching 10 keV. Using processed H$_2$O-ice films, as in \cite{galli_02_2018}, may act to suppress the relative contribution of H$_2$O further, as the total yield can be enhanced temporarily due to the presence of O$_2$ below the sample's surface \citep{meier_sputtering_2020}. 

Besides the compositional dependence on energy ($S_e$), we also see a clear increase in the O$_2$ yield with temperature. For ions, this increase with temperature has been attributed to the ability of radiolytically produced radicals to diffuse and increase production of H$_2$ and O$_2$ near the surface \citep{brown_linear_1980, baragiola_sputtering_2003, teolis_formation_2009}. Our findings support a similar process for electrons, as expected from previous low-energy ($\sim$eV) electron irradiation studies showing that $Y_{H_2O}$ remains constant with irradiation temperature \citep{petrik_electron-stimulated_2005} while $Y_{O_2}$ is relatively constant (but still increases slightly) below $\sim$60 K and increases rapidly as temperature increases above $\sim$60 K \citep{petrik_electron-stimulated_2005, petrik_electron-stimulated_2006, orlando_role_2003}. 

\section{Modeling O$_2$ Sputtering} \label{sec:model}

As noted in the Introduction, between ions and electrons, ions have been the main focus of previous sputtering/exosphere modeling studies \citep{marconi_kinetic_2007, fama_sputtering_2008, teolis_cassini_2010, cassidy_magnetospheric_2013, teolis_water_2017}. The most comprehensive model for predicting O$_2$ sputtering yields for any particle irradiating an icy surface is \cite{teolis_water_2017} which builds off their work in \cite{teolis_cassini_2010}.

\cite{teolis_water_2017} calculates the sputtering yield of O$_2$ as 
\begin{equation}
Y_{O_2}(E, T, \beta)= \frac{\epsilon g_{O_2}^0 x_0}{r_0 \cos \beta} \left[1-\exp \left(-\frac{r_0 \cos \beta}{x_0}\right)\right] \left[1+q_0 \exp \left(-\frac{Q}{k_B T}\right)\right],
\label{Eq:YO2}
\end{equation}
\noindent where $\epsilon$ is the effective particle energy contributing to sputtering (total energy $E=\epsilon$ for electrons), $T$ is the irradiation temperature, $\beta$ is the particle's incident angle, $g_{O_2}^0$ is the surface radiolytic yield of O$_2$ ($Y_{O_2}/E$ when $r_0\cos\beta\ll x_0$), $x_0$ is the optimal depth for O$_2$ production, $r_0\cos\beta$ is the particle's range, $q_0$ is the exponential prefactor for the temperature dependence, $k_B$ is the Boltzmann constant, and $Q$ is the ``activation" energy (noted in \citeauthor{teolis_water_2017} \citeyear{teolis_water_2017} to not have a determined physical significance). \cite{teolis_water_2017} fit Equation \ref{Eq:YO2} to existing laboratory data for ions and determined intrinsic parameter values for $g_{O_2}^0$, $x_0$, $q_0$, and $Q$ (listed in Table \ref{Table:Parameters}). 

In \cite{tribbett_sputtering_2021}, we determined that Equation \ref{Eq:YO2} overestimates O$_2$ production from ions with ranges $r_0 \cos \beta >> x_0$ by as much as an order of magnitude and explored how this could be caused by the assumption in \cite{teolis_water_2017} that energy is deposited uniformly over the ion's range. Upon further investigation, we noticed a mistake in \cite{teolis_water_2017} regarding the angle of incidence for data taken by \cite{bar-nun_ejection_1985}\footnote{The \cite{bar-nun_ejection_1985} paper states that their angle of incidence is 60$^{\circ}$. However, it appears that the ions are incident 60$^{\circ}$ with respect to the surface and therefore 30$^{\circ}$ with respect to the surface normal. For our reasoning, compare the text in paragraph two on page 146 to the diagram in Figure 1, the caption of Figure 4, and the text beginning at line 5 on page 151 stating ``in our experiments the [ion] beam was 30$^{\circ}$ away from perpendicular."} for highly penetrating ions, which could also be contributing to the discrepancy between Equation \ref{Eq:YO2} and experimental data. We are hoping to revisit the effects of these corrections in a future study. 

\subsection{Scaling the Model to Electrons}

\begin{figure}[b]
\centering
\includegraphics[width=0.5\textwidth]{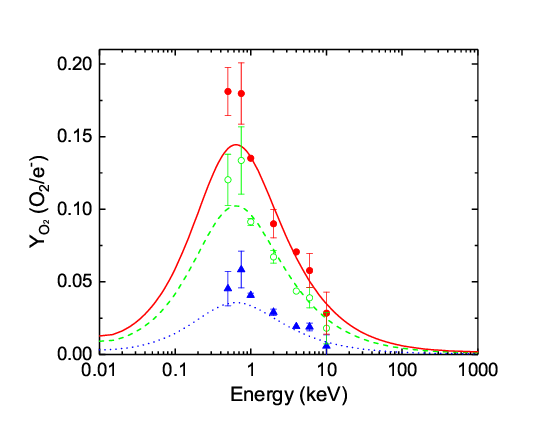}
\caption{Our group's experimental data (0.5 keV from \citeauthor{davis_contribution_2021} \citeyear{davis_contribution_2021}, the rest from this work) for irradiation temperatures of 80 K (\textcolor{blue}{$\blacktriangle$}), 115 K (\textcolor{green}{$\circ$}) and 124.5 K (\textcolor{red}{$\bullet$}) compared to Equation \ref{Eq:YO2} scaled down by our best-fit scaling factor of 0.12 for 80 K (blue dotted line), 115 K (green dashed line), and 124.5 K (red solid line).}
\label{fig:scaling}
\end{figure}

The model's predicted O$_2$ sputtering trends are generally consistent with what has been observed for electrons (see Introduction). Thus, it seems reasonable that first attempts to model electron sputtering simply scale Equation \ref{Eq:YO2}, calculated with parameter values derived using ion data, down by a constant factor ($C*Y_{O_2}$) since at the time there was a lack of electron data with which to determine electron specific parameter values. \cite{teolis_cassini_2010, teolis_water_2017} uses a factor of $C$=0.29 to fit $C*Y_{O_2}$ to experimental O$_2$ yields for low-energy (5 to 30 eV) electrons provided by Petrik, Kavesky, and Kimmel at Pacific Northwest National Laboratory (supplemental Figure S9 in \citeauthor{teolis_cassini_2010} \citeyear{teolis_cassini_2010}), although their measured yields are an order of magnitude higher than the values reported by \cite{sieger_production_1998}\footnote{The data found in the top of Figure 2 in \cite{orlando_role_2003} is the same as in \cite{sieger_production_1998}, but there is a typo in the y-axis label of the plot (Orlando personal communication).} for similar electron energies. 

In two of our recent studies, we applied Equation \ref{Eq:YO2} to our electron sputtering data and found best-fit scaling factors of $C$=0.25 \citep{meier_sputtering_2020} and 0.14 \citep{davis_contribution_2021}, keeping in mind that \cite{meier_sputtering_2020} only measured $Y_T$ and not $Y_{O_2}$. Following this precedent, we find the scaling factor $C$=0.12 minimizes chi-squared between $C*Y_{O_2}$ and all of our group's data listed in Table \ref{Table: Results}. When calculating $Y_{O_2}$, we interpolate our electron ranges from the newly published model predicting the $S_e$ and range of electrons in liquid H$_2$O by \cite{castillo-rico_stopping_2021}, which differs slightly from \cite{grun_lumineszenz-photometrische_1957} and ESTAR \citep{berger_estar_2017} estimates used in \cite{meier_sputtering_2020} and \cite{davis_contribution_2021}, and from estimates by \cite{laverne_penetration_1983} used by \cite{teolis_cassini_2010, teolis_water_2017}. To be consistent with the derived electron ranges in \cite{castillo-rico_stopping_2021}, we assume the density of H$_2$O-ice is the same as liquid H$_2$O (1 g cm$^{-3}$).

As seen in Figure \ref{fig:scaling}, scaling $Y_{O_2}$ down by a constant value results in a reasonable fit above 1 keV for higher temperatures, but underestimates our data at lower energies and lower temperatures, suggesting that the energy and temperature dependencies for ions are not accurately describing trends in all electron data currently available. Thus, as we now have new data for the sputtered component of O$_2$, we reevaluate intrinsic parameter values ($g_{O_2}^0$, $x_0$, $q_0$, and $Q$) in Equation \ref{Eq:YO2} using Markov chain Monte Carlo (MCMC) methods to determine whether we can improve the model's overall fit while removing the need for a constant scaling factor. 

\subsection{Updating Parameter Values for Electrons} \label{subsec:update_model}

Here we present a brief summary of our modeling methods (see Appendix \ref{Appendix: Model} for additional details). Due to the conflicting O$_2$ yields for low-energy ($\sim$10 to 30 eV) electrons \citep{ sieger_production_1998, teolis_cassini_2010}, we excluded both data sets from our initial MCMC analysis. However, we re-ran our MCMC optimization process using a combination of data from our group and \cite{sieger_production_1998} and from our group and \cite{teolis_cassini_2010}. We assume an error of 100$\%$ for data from both \cite{sieger_production_1998} and \cite{teolis_cassini_2010}.
\begin{deluxetable}{c|r@{$\pm$}l r@{$\pm$}l r@{$\pm$}l r@{$\pm$}l}[b]
\tablecaption{Optimized Parameter Values \label{Table:Parameters}}
\tablehead{
\colhead{} & \multicolumn2c{\textbf{$g_{O_2}^0$}} & \multicolumn2c{\textbf{$x_0$}} & \multicolumn2c{\textbf{$q_0$}}  & \multicolumn2c{\textbf{$Q$}} \\
\colhead{\textbf{Data Set}} & \multicolumn2c{(10$^{-4}$ eV$^{-1}$)} & \multicolumn2c{(nm)} & \multicolumn2c{} & \multicolumn2c{(eV)}}
\startdata
Our group only                            & 5.8  & 3 & 3.6  & 1.3 & 960  & 170 & 0.062 & 0.002 \\ 
Our group + \cite{sieger_production_1998} & 2.1  & 1 & 10.3 & 1.0   & 1090 & 220 & 0.064 & 0.002 \\
Our group + \cite{teolis_cassini_2010}    & 7.4  & 2 & 2.8  & 1.0   & 955  & 220 & 0.062 & 0.002 \\
Ions from \cite{teolis_water_2017}        & 50.0 & 5 & 2.8  & 0.4 & 1000 & 100 & 0.06  & 0.01\\ 
\enddata
\end{deluxetable}

Table \ref{Table:Parameters} shows each version of our MCMC optimized values for $g_{O_2}^0$, $x_0$, $q_0$, and $Q$ compared with the values determined in \cite{teolis_water_2017} for ions. Regardless of the electron data set used in the optimization, the resulting $g_{O_2}^0$ value is an order of magnitude smaller than what has been determined for ions. Because $g_{O_2}^0$ is defined as the radiolytic yield at the surface, experimental data for lower energy ($\sim$eV) electrons, which do not travel very deep beneath the surface, heavily influence the optimized $g_{O_2}^0$ value. This explains the variation in $g_{O_2}^0$ values with the three electron data sets, which would likely be larger if we had stricter error for the low-energy data sets. Additionally, $g_{O_2}^0$ and $x_0$ are inversely correlated to each other, which explains $x_0$ increasing when $g_{O_2}^0$ decreases.

\begin{figure}[t]
\centering
\includegraphics[width=0.5\textwidth]{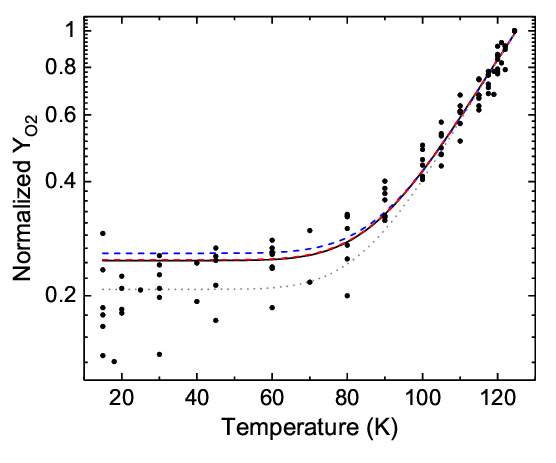}
\caption{Equation \ref{Eq:YO2} calculated using parameter values optimized to data from our group (black solid line), our group and \cite{sieger_production_1998} (blue dashed line), and our group and \cite{teolis_cassini_2010} (red dashed line, mostly overlapping the black solid line) normalized to unity at 124.5 K. Our group's data for every energy normalized to unity at 124.5 K has been included ($\bullet$). For comparison, we also plot Equation \ref{Eq:YO2} using parameter values from \cite{teolis_water_2017} for ions (grey dotted line).}
\label{fig:temp dependence}
\end{figure}

Our optimized values for $q_0$ and $Q$ for the three data sets show less variation than did $g_{O_2}^0$ and $x_0$, and all overlap with each other and with the values obtained from ions within error. As shown in Figure \ref{fig:temp dependence}, the assumption that $Y_{O_2}$ is approximately constant at temperatures $\leq$60 K ignores the observed weak increase in $Y_{O_2}$ at low temperatures, which results in discrepancies between the data and model for temperatures $\lesssim$100 K. Future modeling efforts could potentially modify the structure of Equation \ref{Eq:YO2} to better fit electron data over the entire temperature range.

\begin{figure}[h]
\centering
\includegraphics[width=0.5\textwidth]{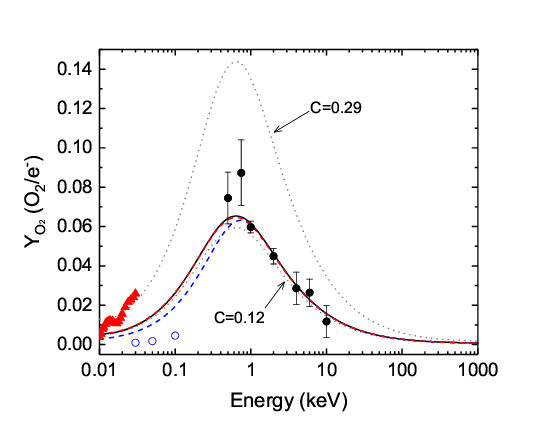}
\caption{Modeled O$_2$ sputtering yields ($Y_{O_2}$) using Equation \ref{Eq:YO2} and ion parameter values \citep{teolis_water_2017} multiplied by our best fit scaling factor $C=0.12$ (grey dotted line, labeled on plot) and $C=0.29$ from \cite{teolis_water_2017} (grey dotted line, labeled on plot), and calculated with parameter values optimized using MCMC methods to match data from our group (black solid line), our group and \cite{sieger_production_1998} (blue dashed line), and our group and \cite{teolis_cassini_2010} (red dashed line, mostly overlapping the black solid line) for an irradiation temperature of 100 K and an electron angle of incidence of 12.5$^{\circ}$. We also show experimental data at 100 K from our group ($\bullet$) and \cite{teolis_cassini_2010} (\textcolor{red}{$\blacktriangle$}), and note that the data from \cite{sieger_production_1998} (\textcolor{blue}{$\circ$}) is at 110 K.}
\label{fig:Yield MCMC fits}
\end{figure}

We plot the energy dependence of the resulting model fits and data at a single representative temperature in Figure \ref{fig:Yield MCMC fits}, showing $Y_{O_2}$ calculated using each set of optimized parameter values for electrons listed in Table \ref{Table:Parameters} compared to $C*Y_{O_2}$ for $C$=0.12 (our best-fit scaling factor) and $C$=0.29 \citep{teolis_cassini_2010,teolis_water_2017} with $Y_{O_2}$ calculated using parameter values for ions from \cite{teolis_water_2017}. The $C$=0.29 fit used by \cite{teolis_cassini_2010, teolis_water_2017} overestimates all of our measured O$_2$ sputtering rates for keV electrons. Conversely, our best fit scaling factor $C$=0.12 underestimates the yields for all of our data (with the exception of 10 keV). Equation \ref{Eq:YO2} optimized to data from only our group and data combined from our group and \cite{teolis_cassini_2010} are very similar (and practically overlap in Figures \ref{fig:temp dependence} and \ref{fig:Yield MCMC fits}), although with better constrained error for data from \cite{teolis_cassini_2010} differences in the resulting curves may be greater.

Finally, we note that the electron energy associated with the peak of $Y_{O_2}$ ($\sim$0.65 keV) does not match the electron energy associated with the peak of electron stopping cross section ($\sim$0.12 keV, \citeauthor{castillo-rico_stopping_2021} \citeyear{castillo-rico_stopping_2021}) even though $Y_T$ is expected to trend with electron stopping cross section. Assuming the peak position of $Y_{O_2}$ should match the peak position of $Y_T$, this difference could be due the overlapping $Y_{O_2}$ values in our 0.5 and 0.75 keV data. Different models also shift the peak position of electron stopping cross section \citep{ashley_simple_1982, laverne_penetration_1983, luo_experimental_1991, gumus_new_2008, castillo-rico_stopping_2021} which could also contribute to this difference, although to a lesser extent.

\section{Astrophysical Implications} \label{sec: AI}

Below, we apply our newly optimized O$_2$ sputtering model for electron irradiation of H$_2$O-ice to three Jovian icy satellites: Europa, Ganymede, and Callisto. For Europa, we compare how the inclusion of low-energy electron experimental sputtering data affects our calculated production yields by calculating Equation \ref{Eq:YO2} with each set of parameter values in Table \ref{Table:Parameters}. When comparing our calculations of Europa to other values in literature, and when discussing Ganymede and Callisto, we calculate Equation \ref{Eq:YO2} with the parameter values optimized to our group's data only.

\subsection{Europa} \label{subsec: Europa}
As done in \cite{davis_contribution_2021}, we calculate the flux of sputtered O$_2$ from Europa as
\begin{equation}
\pi \int J(E) Y_{O_2}(E,T,\beta) d E,
\label{Eq:Integral}
\end{equation}
\noindent where $J(E)$ is the differential flux of electrons (e$^-$ cm$^{-2}$ s$^{-1}$ sr$^{-1}$ MeV$^{-1}$) near Europa, assuming a uniform electron flux striking the surface. We adopt the same differential electron flux that \cite{davis_contribution_2021} estimated by combining measurements from the Galileo Energetic Particle Detector \citep{cooper_energetic_2001} and Voyager Plasma Spectrometer \citep{scudder_survey_1981, sittler_io_1987}.

We integrate Equation \ref{Eq:Integral} between 10 eV (the minimum energy required for electron sputtering; \citeauthor{sieger_production_1998} \citeyear{sieger_production_1998}; \citeauthor{orlando_role_2003} \citeyear{orlando_role_2003}) and 1 MeV, calculating $Y_{O_2}$ with Equation \ref{Eq:YO2}, intrinsic parameter values from Table \ref{Table:Parameters}, and assuming an average $\beta$ of 45$^{\circ}$. Additionally, since \cite{castillo-rico_stopping_2021} only calculate electron ranges up to $\sim$430 keV, we used \cite{castillo-rico_stopping_2021} ranges for energies $\leq$425 keV and scaled ESTAR estimated electron ranges from 425 keV to 1 MeV by a factor of 1.011 so that ranges from \cite{castillo-rico_stopping_2021} and ESTAR matched at 425 keV. 
\begin{deluxetable}{c|cc|cc}[b]
\tablecaption{Global Electron Sputtering from Europa. \label{Table:GlobalRates}}
\tablehead{
\colhead{} & \multicolumn{2}{c}{\textbf{O$_2$ Sputtered Flux}} & \multicolumn{2}{c}{\textbf{O$_2$ Rate}} \\
\colhead{} & \multicolumn{2}{c}{(10$^8$ cm$^{-2}$ s$^{-1}$)} & \multicolumn{2}{c}{(10$^{26}$ s$^{-1}$)} \\
\colhead{\textbf{Parameter Values Used in Equation \ref{Eq:YO2}}} & \colhead{80 K} & \colhead{125 K} & \colhead{80 K} & \colhead{125 K} }
\startdata
Ion, scaled by 0.12   & 1.4 & 5.9 & 0.44 & 1.8 \\ 
Ion, scaled by 0.29   & 3.4 & 14  & 1.1  & 4.4 \\ 
Electron, optimized to data from our group only  & 1.7 & 6.0 & 0.51 & 1.9 \\ 
Electron, optimized to data from our group + \cite{sieger_production_1998} & 1.5 & 5.1 & 0.45 & 1.6 \\
Electron, optimized to data from our group + \cite{teolis_cassini_2010}    & 1.7 & 6.1 & 0.52 & 1.9 
\enddata
\end{deluxetable}

Table \ref{Table:GlobalRates} shows the flux of sputtered O$_2$ and global production rates from Europa found by scaling the sputtered flux to Europa's surface area (mean radius from \citeauthor{showman_galilean_1999} \citeyear{showman_galilean_1999}) for relevant surface temperatures \citep{spencer_temperatures_1999, ashkenazy_surface_2019}. Interestingly, with the exception of scaling $Y_{O_2}$ for ions down by $C$=0.29 which effectively doubles the production rate of O$_2$, the choice of parameter values used to calculate the O$_2$ production rate does not appear to matter significantly. For instance, there is only $\sim$5$\%$ difference between O$_2$ production rates at 125 K found by multiplying $Y_{O_2}$ for ions down by $C$=0.12 (our best fit scaling factor) and calculating $Y_{O_2}$ with parameter values found by optimizing $Y_{O_2}$ to our group's data. The O$_2$ production rates found by calculating $Y_{O_2}$ with parameter values optimized to data from our group and \cite{sieger_production_1998} or data from our group and \cite{teolis_cassini_2010} differ from one another by $\sim$17$\%$. While a $\sim$17$\%$ difference in the O$_2$ production rate for the two low-energy data sets is not seemingly large, as noted in Section \ref{subsec:update_model}, having better constrained error for the low-energy data sets would increase the difference in the resulting integrated yield. Further refining these discrepancies would require additional measurements with low-energy ($\sim$eV) electrons, which would enable a more precise estimate of the O$_2$ surface radiolytic yield ($g_{O_2}^0$).

While we calculated the values in Table \ref{Table:GlobalRates} assuming a uniform electron flux striking the surface of Europa, this is an oversimplification of the radiation environment \citep{paranicas_electron_2001, paranicas_europas_2009, patterson_characterizing_2012, dalton_exogenic_2013, addison_surfaceplasma_2023}. Future studies investigating to what degree spatial variations in electron flux alter our estimates are important for properly applying our optimized electron model to Europa. Regardless, we find a global production rate of (0.5 - 1.9) x 10$^{26}$ O$_2$ s$^{-1}$ for 80 to 125 K using the parameter values optimized to our group's data, which is slightly lower than our previous less-refined estimate \citep{davis_contribution_2021}. Additionally, our estimate for 125 K is a factor of $\sim$1.6 times higher than the estimate made in \cite{vorburger_europas_2018} (1.15 x 10$^{26}$ O$_2$ s$^{-1}$, found by multiplying the sum of the O$_2$ yields from both hot and cold electrons listed in their Table 5 by the surface area of Europa). Considering that, at the time of their study, the only measurement for $Y_{O_2}$ suggested that $Y_{O_2}$ was constant above 200 eV and about an order of magnitude higher than what we have measured at 1 keV \citep{galli_sputtering_2017}, the similarity of the estimates may seem surprising. However, \cite{vorburger_europas_2018} also assumed that only 20$\%$ of the electron flux below 1 keV reaches Europa's surface. Recently, \cite{addison_surfaceplasma_2023} combined the low-energy (thermal) electron sputtering rate estimated in \cite{vorburger_europas_2018} ($\sim$2.3 x 10$^{25}$ O$_2$ s$^{-1}$) with a new sputtering rate estimate for 5 keV to 10 MeV electrons taking into account interactions between Jupiter's magnetosphere and Europa's induced magnetic field, and found the total sputtering contribution from electrons to be only $\sim$2.4 x 10$^{25}$ O$_2$ s$^{-1}$. 

While our assumption that all thermal electrons reach Europa's surface is unlikely, it is also unlikely that there is a constant 80$\%$ reduction in flux for all electron energies below 1 keV \citep{vorburger_europas_2018, addison_surfaceplasma_2023}. Until there is better understanding of what portion of the lower energy ($\lesssim$5 keV) electron flux reaches Europa's surface, we consider our estimates to be an upper limit, as we have suggested previously based on another recent, but lower, flux estimate \citep{jun_updating_2019}. In fact, using fluxes from \cite{jun_updating_2019} results in a rate of (1.0 - 3.7) x 10$^{25}$ O$_2$ s$^{-1}$, which is a factor of 5 lower than our production rate, putting it in range of the value estimated by \cite{addison_surfaceplasma_2023}. This similarity is a bit surprising, considering the estimate from \cite{addison_surfaceplasma_2023} is considerably more refined than ours with the inclusion of spatially resolved energetic electron fluxes, surface temperature differences, and various incident particle angles. 

Our estimated range for the global electron sputtering rate of O$_2$ from Europa encompasses the total production rate for all ions combined of $\sim$1 x 10$^{26}$ O$_2$ s$^{-1}$ estimated in both \cite{cassidy_magnetospheric_2013} and \cite{addison_influence_2021, addison_effect_2022} using the unmodified Equation \ref{Eq:YO2} from \cite{teolis_water_2017}. While the effects of ions and electrons are unlikely to simply be additive, it is interesting that the sum of the estimate for ions and our electron estimate is similar to an estimate of O$_2$ production from Europa's surface via radiation processing ((2.2$\pm$1.2) x 10$^{26}$ O$_2$ s$^{-1}$), which was extrapolated from measurements of atmospheric H$_2$ loss rates during Juno’s recent fly-by of Europa \citep{szalay_oxygen_2024}. Considering the possible reduction of our electron sputtering rate estimates from the deflection of thermal electrons near Europa and that we also recently found Equation \ref{Eq:YO2} likely overestimates $Y_{O_2}$ by a factor of 5 to 8 at 120 K for 0.5 to 5 keV ions \citep{tribbett_sputtering_2021}, which are representative of the cold/thermal ion component near Europa, more rigorous investigation is needed to determine whether the apparent agreement with the Juno-derived data is fortuitous. Nonetheless, we expect that, at the very least, electrons are significant contributors to the sputter-produced O$_2$ exosphere around Europa and need to be considered in any future modeling efforts.

\subsection{Ganymede} \label{subsec: Ganymede}
Ganymede has an exosphere predominately composed of O$_2$, atomic O, and H$_2$O \citep{hall_far-ultraviolet_1998, de_kleer_optical_2023} hypothesized to be produced via sputtering and sublimation. Sputtering from Ganymede by Jupiter's magnetospheric particles is complicated by Ganymede's intrinsic magnetic field deflecting certain energetic particles away from the moon's surface \citep{delitsky_ice_1998, plainaki_h2o_2015, fatemi_formation_2016, poppe_thermal_2018, liuzzo_variability_2020}. A recent study \citep{liuzzo_variability_2020} showed Ganymede's closed field lines around its equator completely shield the moon's equatorial region from irradiating electrons with energies $\lesssim$40 MeV, while electrons of all energies reach the surface of Ganymede's polar regions \citep{frank_lowenergy_1997, cooper_energetic_2001, liuzzo_variability_2020}. 

We estimate the flux of O$_2$ sputtered from Ganymede by electrons with Equation \ref{Eq:Integral}, calculating $Y_{O_2}$ with Equation \ref{Eq:YO2}, parameter values optimized to our group's data, and assuming $J(E)$ for electrons near Ganymede's orbital radius \citep{paranicas_energy_2021} contributes to a uniform electron flux striking Ganymede's polar regions. We extrapolate the differential electron fluxes given in \cite{paranicas_energy_2021} down to 10 eV in order to integrate from 10 eV to 1 MeV. We find the flux of sputtered O$_2$ to be (3 - 20) x 10$^{7}$ cm$^{-2}$ s$^{-1}$ for 65 to 140 K (limits for the temperature range at Ganymede's poles; \citeauthor{squyres_surface_1980} \citeyear{squyres_surface_1980}). Using the mean radius for Ganymede \citep{showman_galilean_1999} and scaling our O$_2$ sputtered flux estimate by the area of the polar regions where electrons reach the surface (latitudes $\geq$40$^{\circ}$; \citeauthor{liuzzo_variability_2020} \citeyear{liuzzo_variability_2020}) yields a production rate of (0.9 - 6.2) x 10$^{25}$ O$_2$ s$^{-1}$ for 65 to 140 K, which, to our knowledge, is the first estimate for O$_2$ production rates from Ganymede by irradiating electrons. While our estimate could be refined further, it appears to be significantly lower than the most recent estimate for O$_2$ production by ions of 2.4 x 10$^{26}$ O$_2$ s$^{-1}$ by \cite{pontoni_simulations_2022}, and even lower than earlier estimates for ions by \cite{marconi_kinetic_2007} (1.2 x 10$^{27}$ O$_2$ s$^{-1}$) and \cite{plainaki_h2o_2015} (2.6 x 10$^{28}$ O$_2$ s$^{-1}$). However, given that \cite{pontoni_simulations_2022} used Equation \ref{Eq:YO2} and found that low-energy ($\sim$keV) ions were the major contributor to the O$_2$ sputtering flux, it is possible that the production rates for ions were overestimated (see above). Regardless, given the simplicity of our estimate, as well as the wide range of estimates for sputtering from ions, more studies investigating sputtering from Ganymede's surface may be merited. 

\subsection{Callisto} \label{subsec: Callisto}
O$_2$ has also been detected in Callisto's atmosphere \citep{cunningham_detection_2015, de_kleer_optical_2023}. While sputtering has been speculated to play a key role for decades \citep{kliore_ionosphere_2002}, surface particle fluxes are difficult to assess due to the presence of Callisto's atmosphere \citep{strobel_italhubble_2002}. Recently, \cite{carberry_mogan_callistos_2023} modeled the spatial variation in temperature and particle (electron and hydrogen, oxygen, and sulfur ion) fluxes across the surface of Callisto. They then calculated sputtering rates for ions using both the model developed by \cite{fama_sputtering_2008} and modified by \cite{johnson_composition_2009} as well Equation \ref{Eq:YO2} \citep{teolis_water_2017} with parameter values for ions modified in \cite{tribbett_sputtering_2021}, and for electrons using Equation \ref{Eq:YO2} with parameter values derived for ions \citep{teolis_water_2017} and no scaling factor. They determine that, although sputtering from Callisto's surface is not enough to account for the observed column densities of O$_2$ in Callisto's atmosphere, electrons contribute between 24 to 32$\%$ of the total O$_2$ sputtered from Callisto's icy patches, which is more than the contribution from hydrogen ($\sim$0.5 to 7$\%$) and oxygen (18 to 20$\%$) ions, but less than from sulfur ions (57 to 41$\%$). 

While recalculating the spatially resolved sputtering yields from Callisto's ice patches is beyond the scope of this work, here we estimate the effect that our new electron parameter values have on the predicted O$_2$ sputtered from Callisto's surface. Although we cannot use Equation \ref{Eq:Integral}, due to the complication of the impinging flux interacting with the moon's atmosphere, we make a rough estimate by integrating the radiolytic yield ($G_{O_2}=Y_{O_2}/E$) at temperatures relevant for Callisto (80 to 144 K; \citeauthor{grundy_near-infrared_1999} \citeyear{grundy_near-infrared_1999}; \citeauthor{carberry_mogan_callistos_2023} \citeyear{carberry_mogan_callistos_2023}) with our new electron parameter values and compare that result with what we obtain when we perform the integration with the parameters values used in \cite{carberry_mogan_callistos_2023}. Integrating from 10 eV to 1 MeV, we find that our new electron parameter values reduce the yield by an order magnitude, suggesting that the contribution of electrons to sputtering of O$_2$ from Callisto is likely significantly less than as estimated by \cite{carberry_mogan_callistos_2023}. 

\section{Conclusions} \label{sec:conclusion}
In this paper, we measured the total, H$_2$O, and O$_2$ sputtering yields for electrons with energies between 0.75 and 10 keV and for irradiation temperatures between 15 and 124.5 K. Over our studied energies, we found that both total and O$_2$ yields increase with decreasing energy (increasing $S_e$) and increase rapidly at temperatures above 60 K, which is in agreement with our previous electron work, as well as previous studies with electrons and ions. In addition, we find that the yield of H$_2$O has a nearly quadratic relation with $S_e$ while the yield of O$_2$ appears to trend approximately linearly with $S_e$ (although the slope changes with temperature). These different dependencies could explain why the trend of $Y_T$ with $S_e$ ranges from quadratic to linear in previous studies for light ions and electrons. Additionally, the composition of the sputtered flux has a strong dependence on electron energy with the relative amount of H$_2$O decreasing rapidly with decreasing $S_e$ over the electron energies studied. In fact, we find that above 4 keV, the contribution from H$_2$O is essentially zero within the limits of our error.

Combining our data with O$_2$ sputtering yields for 0.5 keV electrons from \cite{davis_contribution_2021} and other low-energy ($\sim$eV) electron data from literature \citep{sieger_production_1998, teolis_cassini_2010}, we reevaluated intrinsic parameters in the sputtering model from \cite{teolis_water_2017}, finding that we can provide a more satisfying fit while also removing the arbitrary scaling factor. Having better constraints on low-energy electron O$_2$ yields in literature or restructuring the energy and/or temperature dependent model components may improve the fit further. 

Combining our newly optimized sputtering model with incoming electron fluxes near Europa, we calculate that electrons may contribute to the production of Europa's O$_2$ exosphere at a rate similar to all ion types combined. Thus, although electrons may, in most cases, have significantly lower individual sputtering yields than ions, the higher electron fluxes at the surface of icy bodies like Europa may be large enough for electrons to be a major contributor to exospheric O$_2$ production. In contrast, we find electrons contribute less to O$_2$ sputtering from Ganymede and Callisto, although the contribution of electrons is still likely non-negligible. Of course, future studies for all moons examining spatial variations in the incoming electron flux are needed to refine our estimates. Regardless, at the very least, it seems clear that the contribution of electrons needs to be included in sputtering and exosphere modeling of icy bodies going forward.

\vspace{\baselineskip}
\noindent We would like to thank T.M. Orlando and B.D. Teolis for explaining their group's published electron sputtering data as well as P.D. Tribbett for contributing to the MCMC methods discussion. This research was supported by NASA Solar System Workings Award $\#$80NSSC20K0464. Data can be found in Northern Arizona University’s long-term repository \url{https://openknowledge.nau.edu/id/eprint/6258}).

\begin{appendix}
\section{Data} \label{Appendix: Data}
In Table \ref{Table: Results}, we list our measured total, H$_2$O, and O$_2$ sputtering yields for each electron energy and temperature studied in this paper and in \cite{davis_contribution_2021}. Molecular H$_2$ sputtering yields are calculated as $Y_{H_2}=2*Y_{O_2}$.

\startlongtable
\begin{deluxetable*}{c|DDD}
\tablecaption{All Laboratory Sputtering Yields \label{Table: Results}}
\tabletypesize{\small}
\tablehead{
\colhead{} & \multicolumn2c{\textbf{Total}} & \multicolumn2c{\textbf{H$_2$O}} & \multicolumn2c{\textbf{O$_2$}} \\
\colhead{\textbf{Temperature}} & \multicolumn2c{\textbf{Yield}} & \multicolumn2c{\textbf{Yield}} & \multicolumn2c{\textbf{Yield\tablenotemark{a}}} \\
\colhead{(K)} & \multicolumn2c{(10$^{-24}$ g/e$^-$)} & \multicolumn2c{(H$_2$O/e$^-$)} & \multicolumn2c{(O$_2$/e$^-$)}}
\decimals
\startdata
\multicolumn7c{\textbf{0.5 keV electrons}\tablenotemark{b}} \\ \hline
15    & 6.62  & 0.17   & 0.025 \\
30    & 6.63  & 0.17   & 0.025 \\
45    & 6.98  & 0.17   & 0.031 \\
60    & 7.13  & 0.17   & 0.034 \\
80    & 7.83  & 0.17   & 0.045 \\
90    & 8.60  & 0.17   & 0.058 \\
100   & 9.57  & 0.17   & 0.075 \\
105   & 10.23 & 0.17   & 0.086 \\
110   & 11.77 & 0.17   & 0.111 \\
115   & 12.31 & 0.17   & 0.120 \\
117.5 & 13.58 & 0.17   & 0.142 \\
120   & 13.56 & 0.17   & 0.141 \\
122   & 14.80 & 0.17   & 0.162 \\
124.5 & 15.94 & 0.17   & 0.181 \\ \hline
\multicolumn7c{\textbf{0.75 keV electrons}} \\ \hline
15    & 5.32   & 0.094  & 0.042 \\
20    & 5.05   & 0.094  & 0.038 \\
30    & 5.39   & 0.094  & 0.043 \\
45    & 5.46   & 0.094  & 0.045 \\
60    & 5.57   & 0.094  & 0.046 \\
80    & 6.29   & 0.094  & 0.058 \\
90    & 7.19   & 0.094  & 0.072 \\
100   & 8.02   & 0.094  & 0.087 \\
105   & 8.97   & 0.094  & 0.103 \\
110   & 9.62   & 0.094  & 0.114 \\
115   & 10.79  & 0.094  & 0.134 \\
117.5 & 11.03  & 0.094  & 0.138 \\
120   & 11.84  & 0.094  & 0.151 \\
122   & 12.61  & 0.094  & 0.164 \\
124.5 & 13.55  & 0.094  & 0.180 \\ \hline
\multicolumn7c{\textbf{1 keV electrons}} \\ \hline
15    & 3.54  & 0.068  & 0.025 \\
20    & 3.53  & 0.068  & 0.025 \\
30    & 3.74  & 0.068  & 0.027 \\
45    & 4.19  & 0.068  & 0.036 \\
60    & 3.96  & 0.068  & 0.032 \\
80    & 4.46  & 0.068  & 0.041 \\
90    & 5.04  & 0.068  & 0.050 \\
100   & 5.61  & 0.068  & 0.060 \\
105   & 6.30  & 0.068  & 0.071 \\
110   & 6.63  & 0.068  & 0.077 \\
115   & 7.49  & 0.068  & 0.091 \\
117.5 & 7.87  & 0.068  & 0.098 \\
120   & 8.44  & 0.068  & 0.107 \\
122   & 9.39  & 0.068  & 0.123 \\
124.5 & 10.11 & 0.068  & 0.135 \\ \hline
\multicolumn7c{\textbf{2 keV electrons}} \\ \hline
15    & 2.05  & 0.016  & 0.026 \\
30    & 1.86  & 0.016  & 0.023 \\
40    & 1.79  & 0.016  & 0.022 \\
60    & 1.99  & 0.016  & 0.025 \\
70    & 2.08  & 0.016  & 0.027 \\
80    & 2.22  & 0.016  & 0.029 \\
90    & 2.55  & 0.016  & 0.035 \\
100   & 3.17  & 0.016  & 0.045 \\
105   & 3.36  & 0.016  & 0.048 \\
110   & 3.75  & 0.016  & 0.055 \\
115   & 4.50  & 0.016  & 0.067 \\
120   & 5.10  & 0.016  & 0.077 \\
124.5 & 5.86  & 0.016  & 0.090  \\ \hline
\multicolumn7c{\textbf{4 keV electrons}} \\ \hline
15    & 0.93  & 0.006 & 0.013 \\
18    & 0.74  & 0.006 & 0.009 \\
25    & 1.05  & 0.006 & 0.015 \\
30    & 1.06  & 0.006 & 0.015 \\
45    & 1.08  & 0.006 & 0.015 \\
60    & 1.17  & 0.006 & 0.017 \\
80    & 1.32  & 0.006 & 0.019 \\
90    & 1.69  & 0.006 & 0.025 \\
100   & 1.89  & 0.006 & 0.029 \\
105   & 2.03  & 0.006 & 0.031 \\
110   & 2.34  & 0.006 & 0.036 \\
115   & 2.79  & 0.006 & 0.044 \\
117.5 & 3.06  & 0.006 & 0.048 \\
119   & 3.04  & 0.006 & 0.048 \\
120   & 3.42  & 0.006 & 0.054 \\
121   & 3.65  & 0.006 & 0.058 \\
122   & 3.51  & 0.006 & 0.056 \\
124.5 & 4.40  & 0.006 & 0.071 \\ \hline
\multicolumn7c{\textbf{6 keV electrons}} \\ \hline
15    & 0.57  & 0      & 0.010 \\ 
20    & 0.62  & 0      & 0.010 \\
30    & 0.78  & 0      & 0.013 \\
45    & 0.88  & 0      & 0.015 \\
60    & 0.90  & 0      & 0.015 \\
80    & 1.13  & 0      & 0.019 \\
90    & 1.12  & 0      & 0.019 \\
100   & 1.58  & 0      & 0.026 \\
105   & 1.62  & 0      & 0.027 \\
110   & 2.12  & 0      & 0.035 \\
115   & 2.33  & 0      & 0.039 \\
120   & 2.99  & 0      & 0.050 \\
122   & 3.15  & 0      & 0.053 \\
124.5 & 3.45  & 0      & 0.058 \\ \hline
\multicolumn7c{\textbf{10 keV electrons}} \\ \hline
20    & 0.50  & 0.004 & 0.006 \\
40    & 0.45  & 0.004 & 0.005 \\
60    & 0.58  & 0.004 & 0.008 \\
70    & 0.49  & 0.004 & 0.006 \\
80    & 0.46  & 0.004 & 0.006 \\
90    & 0.66  & 0.004 & 0.009 \\
100   & 0.82  & 0.004 & 0.012 \\
105   & 0.96  & 0.004 & 0.014 \\
110   & 1.27  & 0.004 & 0.019 \\
115   & 1.20  & 0.004 & 0.018 \\
117.5 & 1.42  & 0.004 & 0.022 \\
119   & 1.45  & 0.004 & 0.022 \\
120   & 1.67  & 0.004 & 0.026 \\
121   & 1.71  & 0.004 & 0.026 \\
124.5 & 1.82  & 0.004 & 0.028 \\
\enddata
\tablenotetext{a}{molecular yield of H$_2$ is twice that of O$_2$ ($Y_{H_2}=2*Y_{O_2}$)}
\tablenotetext{b}{from \cite{davis_contribution_2021}}
\end{deluxetable*}

\section{Model Optimization} \label{Appendix: Model}

To determine the values of $g_{O_2}^0$, $x_0$, $q_0$ and $Q$ in Equation \ref{Eq:YO2}, \cite{teolis_cassini_2010, teolis_water_2017} split the equation into separate energy and temperature dependencies. They fit the temperature-independent data ($\leq$80 K) to the energy-dependent component to determine $g_{O_2}^0$ and $x_0$ and then found $q_0$ and $Q$ by fitting the temperature-dependent component to O$_2$ yields for all energies normalized to unity at 150 K. Our use of MCMC methods to optimize the model to electron data enables us to determine values for all parameters ($g_{O_2}^0$, $x_0$, $q_0$, and $Q$) without splitting Equation \ref{Eq:YO2} into energy and temperature components. This allows us to optimize the model to every data point available regardless of energy or temperature.

We use ``emcee," an open-source software package in Python \citep{foreman-mackey_emcee_2013} to optimize Equation \ref{Eq:YO2} from \cite{teolis_water_2017} to electron laboratory data. While more commonly used for observational astronomy (\citeauthor{dunkley_fast_2005} \citeyear{dunkley_fast_2005}; \citeauthor{line_uniform_2015} \citeyear{line_uniform_2015}; \citeauthor{tribbett_titan_2021} \citeyear{tribbett_titan_2021}; and many others), our group has previously used emcee to match existing models/equations to experimental data \citep{behr_compaction_2020, carmack_pore_2023}.

Briefly, emcee uses Markov chain Monte Carlo (MCMC) methods with Bayesian inference (as described in \citeauthor{foreman-mackey_emcee_2013} \citeyear{foreman-mackey_emcee_2013} and \citeauthor{behr_compaction_2020} \citeyear{behr_compaction_2020}) in order to explore the probability distribution of parameters in a model when compared to an observed data set. It does this by utilizing ``walkers" which move around the model's multi-dimensional parameter space along a Markov chain. Each ``step" in the Markov chain, or change in parameter values, depends only on how the probability of the current walker values compare to a random sampling of possible new values. With enough steps in the chain, MCMC forgets the user-specified initial parameter values and is able to escape local solutions. Moreover, no additional knowledge (other than defining the likelihood function) is needed to run emcee, eliminating the constraints of grid searches such as user defined spacing/resolution and limiting values. Furthermore, while the points in a grid search scale exponentially with dimensionality, this is not necessarily the case with MCMC, potentially making computational times with MCMC faster. This puts MCMC above other commonly used fitting methods (e.g. by eye or using a grid method, see \citeauthor{speagle_conceptual_2019} \citeyear{speagle_conceptual_2019} for more details) by thoroughly exploring the probability of observed data being described by the model throughout the multi-dimensional parameter space.

We gave all parameters ($g_{O_2}^0$, $x_0$, $q_0$, and $Q$) flat priors limiting them to physical values (i.e. $\geq$0), and we gave $x_0$ an additional Gaussian prior of 5 $\pm$ 4 nm in order to encompass estimates of efficient O$_2$ production depths from \cite{petrik_electron-stimulated_2006} and \cite{meier_sputtering_2020}. Walkers in MCMC can and will stray away from the mean of the specified Gaussian prior if the likelihood of the observed data given the model prefers it. We randomly distributed initial parameter values for walkers around an estimate made by fitting Equation \ref{Eq:YO2} to the data by eye. While not necessary, starting walkers with an educated guess will reduce the number of steps and therefore computational time the walkers need in order to constrain the posterior distribution and best fit parameter values (unless of course the educated guess was a poor one or a local solution).

We ran our MCMC optimization process with three $Y_{O_2}$ data sets: from our group, from our group and \cite{sieger_production_1998}, and from our group and \cite{teolis_cassini_2010}. We modeled the data from \cite{sieger_production_1998} and \cite{teolis_cassini_2010} separately because of their large differences in $Y_{O_2}$ for similar electron energies (see Figure \ref{fig:Yield MCMC fits}). We scaled their data by $cos^{1.3}(\beta)/cos^{1.3}(12.5^{\circ})$ \citep{vidal_angular_2005} to account for the differences in incidence angle ($\beta$) and assumed 100$\%$ error for data from both \cite{sieger_production_1998} and \cite{teolis_cassini_2010} due to no error being provided in the original manuscripts and due to possible differences in sample thickness affecting the reported sputtering yields \citep{petrik_electron-stimulated_2005}. Like we did for our own data, we interpolated electron ranges from \cite{castillo-rico_stopping_2021} for the relevant electron energies used by \cite{sieger_production_1998} and \cite{teolis_cassini_2010}. We list the optimized values for $g_{O_2}^0$, $x_0$, $q_0$, and $Q$ for all data sets in Table \ref{Table:Parameters} and provide additional discussion in Section \ref{subsec:update_model}.

\end{appendix}

\bibliography{maintext.bib}
\bibliographystyle{aasjournal}

\end{document}